\documentclass[superscriptaddress,preprint,nofootinbib, showpacs]{revtex4-1}

\usepackage{amsmath,amsfonts,amssymb,graphicx,color,
times,bbm,psfrag,dsfont}
\usepackage{bbold}
\usepackage{mathrsfs}
\usepackage{slashed}
\usepackage[centerlast]{caption}
\usepackage{float}
\usepackage{subfig}
\usepackage{tikz}
\usepackage{pgfplots}
\usepackage[theorems,skins]{tcolorbox}
\newtcolorbox{mymathbox}[1][]{colback=white, ams gather, outer arc=0pt, #1}
\usepackage{mathrsfs}
\usepackage{epstopdf}
\usepackage{multirow}
\usepackage{gensymb} 
\usepackage{ulem} 
\usepackage[colorlinks=true, citecolor=blue, linkcolor= blue, urlcolor=blue,bookmarks=true]{hyperref}
\usepackage{xr}
\usepackage{zref-xr}
\usepackage{chngcntr}

\def \beq{\begin{equation}}
\def \eeq{\end{equation}}
\def \bse{\begin{subequations}}
\def \ese{\end{subequations}}
\def \bea{\begin{eqnarray}}
\def \eea{\end{eqnarray}}
\def \bem{\begin{displaymath}}
\def \eem{\end{displaymath}}
\def \bem{\begin{pmatrix}}
\def \eem{\end{pmatrix}}

\def \bb{\bibitem}
\def \bs{\boldsymbol}
\def \nn{\nonumber}

\newcommand{\expect}[1]{\langle#1\rangle}

\newcommand{\ads}[1]{\text{AdS}$_#1$}

\newcommand{\Li}[1]{\text{Li}_2 \left( #1 \right)}

\begin{document}

\title{Log-rise of the resistivity in the holographic Kondo Model}
\author{Bikash Padhi}
\email{bpadhi2@illinois.edu}
\affiliation{Department of Physics and Institute for Condensed Matter Theory, University of Illinois, 1110 W. Green Street, Urbana, IL 61801, USA}
\author{Apoorv Tiwari}
\affiliation{Department of Physics and Institute for Condensed Matter Theory, University of Illinois, 1110 W. Green Street, Urbana, IL 61801, USA}
\affiliation{Perimeter Institute for Theoretical Physics, Waterloo, ON, Canada}
\author{Chandan Setty} 
\author{Philip W. Phillips}
\affiliation{Department of Physics and Institute for Condensed Matter Theory, University of Illinois, 1110 W. Green Street, Urbana, IL 61801, USA}

\begin{abstract} 
We study a single-channel Kondo effect using a recently developed  \cite{HoloKondo,HoloKondo2pt, HoloKondoFano, HoloKondo2imp} holographic large-$N$ technique. In order to obtain resistivity of this model, we introduce a probe field. The gravity dual of a localized fermionic impurity in 1+1-dimensional host matter is constructed by embedding a localized 2-dimensional Anti-de Sitter (\ads{2})-brane in the bulk of \ads{3}. This helps us construct an impurity charge density which acts as a source to the bulk equation of motion of the probe gauge field. The functional form of the charge density is obtained independently by solving the equations of motion for the fields confined to the \ads{2}-brane. The asymptotic solution of the probe field is dictated by the impurity charge density, which in turn, affects the current-current correlation functions, and hence the resistivity. Our choice of parameters tunes the near-boundary impurity current to be marginal, resulting in a $\log T$ behavior in the UV resistivity, as is expected for the Kondo problem.  The resistivity at the IR fixed point turns out to be zero, signaling a complete screening of the impurity. 
\end{abstract}


\maketitle

\section{Introduction}
It is truly fortuitous that the word Kondo in Swahili means battle or war because the physics behind the Kondo effect \cite{BlueBook} is reminiscent of just that.  Namely, below some crossover temperature, a lone magnetic impurity in a sea of noninteracting electrons is robbed of its spin as a result of the screening cloud the conduction electrons form around the spin.  The resultant bound state \cite{Nozierres, weigman,  andrei} is an example of an emergent low-energy degree of freedom, totally absent from the UV-complete model. The physics of this bound state formation is captured by the renormalization group treatment of this problem.  At high energies, the magnetic impurity and the conduction electrons are independent.  As the high-energy degrees of freedom are integrated out \cite{WilsonRG, andersonyuval}, the exchange interaction between the conduction electrons and the magnetic spins increases logarithmically.   It is the log-divergence of the Kondo exchange coupling that entails the formation of the singlet ground state at $T=0$.  Consequently, at high energies the system is asymptotically free.  A key signature of the logarithmic divergence of the coupling constant is the experimentally observed log-rise of the resistivity \cite{KondoExp}.   

Since the Kondo model only involves a single magnetic impurity, it is the simplest system that exhibits the physics of strong coupling, exemplified by the formation of new degrees of freedom at low energies.  More modern techniques \cite{Affleck95} have shown that the single-impurity nature of this problem makes it amenable to a reformulation as a problem in boundary conformal field theory.  This development coupled with the inherent strong-coupling physics of this model ultimately suggest that this problem is tailor made to be solved by the gauge-gravity duality (holography).  In principle, the Kondo model could be used as a testing ground for the applicability of holography to condensed matter as this problem has an exact solution \cite{weigman,andrei}.  Thus far, encoding Kondo physics within the gauge-gravity setup \cite{HoloKondo,HoloKondo2pt, HoloKondoFano, HoloKondo2imp} has led to two key results which suggests that perhaps much more of Kondo physics can be extracted within this technique:  1) the emergence of a dynamical scale below which the Kondo coupling diverges and 2) power-law scalings of the  IR resistivity and entropy.  The exponent is governed by the dimension of the irrelevant operator that flows the theory away from the IR fixed point.  Thus far, logarithmic scaling of the resistivity has not been reproduced, since the previous models have been based on Chern-Simons (CS) fields which do not have any propagating degrees of freedom. Here we modify these proposals with the inclusion of a Maxwell field, which gives rise to a current at the boundary.

In this work, we construct the current explicitly at the boundary and compute the corresponding resistivity.  A key difference in our work relative to the previous holographic setup is that we explicitly include a boundary chemical potential for the bulk  $U(1)$ gauge field.    This allows us to include explicitly the charge density degrees of freedom in the Kondo model.  It is the resistivity computed from the correlation function of the current operators dual to the bulk $U(1)$ gauge field that gives rise to the $\ln T$ behavior.    

As a result of the s-wave symmetry of the magnetic impurity, the Kondo model can be described as an effective 1+1-dimensional model described by \cite{Affleck95, GabyLargeN}, 
\beq
H = H_F + H_K = \psi^\dag_L i \partial_x \psi^{ }_L \, + \lambda_K \delta(x) \, \vec{S}_{imp} \cdot \vec{S}_{el} \, .
\eeq
Here $\psi^\dag_L$ creates a relativistic left-moving (chiral) free electron. Interaction of the electron spin current, $\vec{S}_{el}=\psi^\dag_L \vec{T} \psi_L^{ }$, and the impurity spin current, $\vec{S}_{imp} = \chi^\dag \vec{T} \chi$, is localized at $x=0$, with strength $\lambda_K \rightarrow 0$. Here, the $\chi$'s are the Abrikosov fermions.  We will consider a large-$N$ model \cite{BickersRMP} in which $N \rightarrow \infty$ but $\lambda_K N$ is assumed fixed \cite{GabyLargeN}. Here, the components of $\vec{T}$ are the $N^2-1$ generators of the spin group $SU(N)$ in the fundamental representation requires the impurity charge to be $\chi^\dag \chi = \mathfrak{q}_i $ and constrains  the physical space. Notice the engineering dimension of $H_K$ is 2; thus, the model essentially describes a classically marginal deformation to a chiral CFT, describing the free electrons. 

Here, we review the holographic construction for \cite{HoloKondo} the Kondo model.  The Kondo CFT (the above Hamiltonian) is invariant under a spin $SU(N)$ and charge $U(1)$ Kac-Moody current algebras. Since the boundary current is an $SU(N)$ singlet, there is no $SU(N)$ bulk dual field.  The $U(1)$ charge current, however, must be described by a bulk gauge field, $A_\mu$, and its associated field strength, $F$, in \ads{3}. Describing multiple flavors of electrons or a $k$-channel Kondo model requires an $SU(k)$ current algebra. The bulk dual of this is a level-$k$ CS field. The Abrikosov fermions also have a $U(1)$ charge symmetry, and  hence this $U(1)$ current must be described by a bulk dual gauge field, $a_m$, and associated field strength, $f$, in \ads{2}.  Defining a scalar operator $\mathcal{O} = \psi^\dag \chi$, in the large-$N$ limit, one can describe the Kondo coupling as a "double-trace" marginal deformation, $\int dx^2 \, \mathcal{O}^\dag \mathcal{O}$. Since $\mathcal{O}$ is charged under the $U(1)$ symmetries of both $\psi$ and $\chi$ fermions, it should be described by a bifundamental scalar bulk dual $\Phi$ with a covariant derivative, $D_m \equiv \partial_m + i q_3 A_m - i q_2 a_m $. These operators localized to the impurity are functions of time only.  Hence, their bulk dual fields must live only on the \ads{2}-brane, embedded in the \ads{3} bulk. We use $m$ and $n$ as indices on the \ads{2} fields and $\mu$ and $\nu$ for the indices on the \ads{3} fields. While it is customary to set $q_3=-q_2=1$ \cite{HoloKondo}, we will work explicitly in the limit $q_3\ll q_2$.  This limit is appropriate as the impurity sits in \ads{2} making it physical to assume that  $q_2>q_3$.  The Kondo action describing this is
\begin{subequations}
\begin{alignat}{3}
& S_{K} = -N \int \sqrt{-g_{(3)}}\, d^3x \left( \mathcal{L}_{\text{AdS}_{3}} +  \delta(x) \mathcal{L}_{\text{AdS}_2} \right)  \, , 
\\
& \mathcal{L}_{\text{AdS}_3} =  
\frac{1}{2\kappa_3^2} \left( R + \frac{2}{L^2}\right) -	\frac{1}{4e_3^2} \, F^2  \, , 
\\ 
& \mathcal{L}_{\text{AdS}_2} =  \frac{1}{4e_2^2} f^2 + (D^m \Phi )^\dag (D_m \Phi) + M^2 \Phi^\dag \Phi  \, .
\end{alignat}
\label{eq:Action}
\end{subequations}
Since the bulk spacetime is $2+1$ dimensional, it is natural to consider $U(1)$ CS theory in addition to Maxwell theory \cite{MCSbasics}. However, we shall confine our discussion to the UV limit of the Maxwell coupling, $e^2 \ll 1$, in which case the MCS theory flows into a Maxwell theory, which we will use as a probe theory. The gravitational coupling is also taken to be small, $\kappa_3^2 \ll 2e^2$. These limits simplify our calculations enormously. Owing to the probe limit, we work with a fixed, unbackreacted, neutral  background,
\bea
ds^2 = \frac{L^2}{z^2} \left( - h(z) dt^2 + dx^2 + \frac{dz^2}{h(z)}  \right) \, , \\
h(z) = 1 - z^2/z^2_h \quad , \quad T=1/(2 \pi z_h) \, .
\eea
This \ads{3} black hole background is the well-known the Ba\~nados-Teitelboim-Zanelli (BTZ) background. Here, $T$ is the Hawking temperature of the black hole. From now onward, we scale all the lengths by $z_h$ and hence $ T = 1/(2 \pi)$. The \ads{3} radius $L$ is set to unity. The \ads{2}-brane is localized at $x=0$ and has an induced metric $g_{(2)}$; the boundary field theory is located at a fixed inverse-radial cutoff, $z=\epsilon$, and has induced metric $\gamma$. $g_{(3)}$ is the bulk \ads{3}  metric.

\section{Current-current correlation function }
 We wish to compute the dc resistivity of the above theory, which in linear response  can be obtained from the Kubo formula \cite{Hartnoll_Lec}
\beq
\rho_{dc} = \lim_{\bs k \rightarrow 0} \,\, {i \omega} \, \left[ \text{Im } G^R_{xx} (\bs k) \right]^{-1} = \frac{\expect{J_x} }{E_x} \, ,
\label{eq:ResistivityForm}
\eeq
where $G^R_{xx} = \expect{J_x J_x}$ is the retarded current-current correlation function, which is obtained from the on-shell action of Eq. \eqref{eq:Action}. We have denoted $\bs k = (\omega, k)$, and in computing the Green function, we will scale them by $2 \pi T$. The last equality is simply Ohm's law. Using the above expression, we obtain the resistivity in the weak-coupling limit \cite{myNote1}  by evaluating the on-shell action near the boundary of \ads{3}. For obtaining the on-shell action (with a gauge choice $a_z = 0 = A_z $), we first solve the equations of motion (EOM), 
\begin{gather}
\partial_z \left( \sqrt{-g_{(2)}} \, g^{zz} g^{tt} \, \partial_z a_t \right) = - q_2 e_2^2 j^t 
\, , \\
 \partial_z \left( \sqrt{-g_{(2)}} \, g^{zz} \, \partial_z \phi \right) =     \sqrt{-g_{(2)}}  \left(  g^{tt} (q_3 A_t - q_2 a_t )^2  + M^2 \right) \phi 
\, , \\
\partial_z \left(  \sqrt{-g_{(3)}} \, F^{\mu \nu} \right)  =  q_3 e_3^2   \delta(x) {\delta^\nu}_t j^t
\, , \\
2 \sqrt{-g_{(2)}} \, g^{tt} \left(  q_3 A_t - q_2 a_t  \right) \phi = j^t
\, .
\end{gather}
We simplify these equations by appropriately tuning the charges of the scalar field and the Maxwell couplings, such that $q_3\ll q_2 = 1$ and $q_2 e_2^2 \ll q_3 e_3^2=1 $. These two limits allow us to work under two kinds of probe limits. The first limit lets us treat $A_\mu$ as a probe field for the \ads{2}-fields. So, we look for the solutions of $\phi(z)$ and $a_t(z)$, independent of $A_\mu$. The second limit allows us to solve for $a_t$ by setting its source to zero. These limits allow us to work in a scenario in which the dynamics of the impurity fields is not affected by that of the bulk probe field. In the boundary theory this corresponds to the limit when the impurity spin is frozen to the origin and the electronic current interacts with it without altering its dynamics. This helps us simplify the equations to,
\begin{gather}
\partial_z \left( \sqrt{-g_{(2)}} \, g^{zz} g^{tt} \, \partial_z a_t \right) = 0 
\label{eq:EOMat} \, , \\
\partial_z \left( \sqrt{-g_{(2)}} \, g^{zz} \, \partial_z \phi \right) = \sqrt{-g_{(2)}}  \left(  g^{tt} a_t^2 + M^2 \right) \phi
\label{eq:EOMPhi} \, , \\
\partial_\mu \left( \sqrt{-g_{(3)}} \, F^{\mu \nu} \right)  = \delta(x) {\delta^\nu}_t j^t 
\label{eq:EOMAmu} \, , \\
j^t = \sqrt{-g_{(2)}} g^{tt} \, j_0(z) \quad , \quad  j_0(z)  = - 2 a_t \phi^2 .
\end{gather}
First we solve $a_t(z)$ from Eq. \eqref{eq:EOMat},
\beq
a_t(z)  = \mathcal{Q}_i/z  + \mu_i  \, ,
\eeq
Here the charge of the impurity $\mathfrak{q}_i = N \mathcal{Q}_i $ is fixed (Neumann boundary condition) and $N \mu_i$ is the associated chemical potential. Note that $\mathfrak{q}_i, \mu_i$ are emerging parameters of the boundary field theory (fixed by the boundary conditions), which are independent of the choice of the bulk parameters $q_2, q_3$. Now we use this solution for solving $\phi(z)$, where $a_t$ simply modifies the effective mass of the scalar by $O(\mathcal{Q}_i^2)$. We further impose a constraint on the choice of $\mathcal{Q}_i$, $\mathcal{Q}_i^2 - M^2 = 1/4$, such that the Breitenlohner-Freedman stability bound is saturated. This makes the asymptotic solutions of Eq. \eqref{eq:EOMPhi} marginal,
\bea
\phi(z) \simeq z^{1/2} \left( \alpha \log \Lambda z + \beta \right) \,\, , \,\, \alpha = \kappa \beta \, .
\label{eq:SolnPhi}
\eea
 By computing the free energy of the above solution, one can see that the solution of $\phi(z)$ in Eq. \eqref{eq:SolnPhi} is stable only when $T <  - \mu_i/(2 \pi \mathcal{Q}_i) \equiv T_c  \simeq T_K$. For the temperature range $T \geq T_c$, $\phi(z) =0$ turns out to be a more stable solution. Hence, the impurity charge $j^t$ vanishes for $T \geq T_c$, and so does the correlation function. Thus, the impurity (and hence the Kondo effect) can be viewed as a $0+1$-dimensional holographic superconductor, driven by a double-trace marginal coupling $\kappa$ \cite{Faulkner-Multi}. This is expected in large-$N$ model of the Kondo effect \cite{AndreiLargeN}. It is clear form the on-shell action that this condensate $\expect{\mathcal{O}} \propto N \alpha $. For future simplification, we arbitrarily fix the proportionality constant and write, $2 N \alpha = \sqrt{\pi} \expect{\mathcal{O}}$. As pointed out in Ref. \citep{HoloKondo}, $\expect{\mathcal{O}}$ can be thought of as the size of the screening cloud, the formation of which begins below $T<T_c$. Our probe limits ensure a small screening cloud; in other words, our method works only in the temperature range $T \lesssim T_c$. Hence, we asymptotically evaluate $j_0(z)$ only up to the leading order in $\alpha$ or $\expect{\mathcal{O}}$,
\beq
j_0(z) \simeq 2 \mathcal{Q}_i \alpha^2 (z-1)(\ln z + 1/\kappa)^2 \sim O(\alpha^2) \, .
\eeq

Before we proceed to solve the probe fields for the above impurity current, we note that  going beyond the probe limit of $\phi$ essentially corrects the effective mass of $a_t$ by $O(\alpha^2)$. Hence, $j_0$ is corrected to the order  $O(\alpha^4)$, and (as we will see) so does the resistivity. In our limit, this can be safely ignored; however, as $\lambda_k$ grows, one has to include such a correction. In fact, one might also need to revoke the probe limit of $A_\mu$. Beyond the probe limit, the effective mass of $A_\mu$ acquires a correction of  $ O(q_3 \alpha^2 )$. 

In order to probe this current, we introduce Maxwell fields in the bulk. By treating them as linear fluctuations, $A_\mu e^{i (\omega t - k x)}$, we can simplify Eq. \eqref{eq:EOMAmu}. In Lorentzian holography \cite{LorentzHolo}, one can construct the correlation functions from the asymptotic solution of a gauge invariant quantity, such as the electric field $E_x = \omega A_x + k A_t$. Applying infalling  conditions at the horizon and Neumann boundary conditions at the boundary, we solve $E_x$. The leading-order coefficient also provides a spectrum of black hole quasinormal modes, which can be used to construct retarded hydrodynamic correlators \cite{Kovtun}. The EOM of $E_x(z)$ is
\beq
E_x'' + \left( \frac{\omega^2 h'}{ p^2 h } + \frac{1}{z}  \right) E_x' + \frac{p^2}{h^2} E_x = \frac{k j_0}{i z h} \, .
\label{eq:MaxEOM}
\eeq
Here prime denotes the $z$-derivative and $p^2(z) = \omega^2 - h(z) k^2$.  We made a gauge choice, $A_z=0$. Note the source term is proportional to $k$ since it breaks translational invariance. We first solve the above equation without the source $(T > T_c)$ and later build the full solution on top of it. In three dimensions, the asymptotic solution is expected to be of the form $E_x \sim \mathcal{A}_x \ln z + \mathcal{B}_x $, where the coefficients can be fixed by solving the EOM. Using the hydrodynamic expansion \cite{DTSonHydro} in the limit $\omega, k \ll 1$, one can obtain an analytic solution,
\beq
E_x(z) \sim (k^2 - \omega^2) \ln z + (i \omega/2 + \omega^2/4) \, .
\label{eq:ExAss}
\eeq
The derivation of $E_x$ is sketched in Appendix \ref{sec:HydroE}. The coefficients $\mathcal{A}_x$ and $\mathcal{B}_x$ can be identified from the above expression. From these coefficients, one can construct the Green functions as
\beq
\frac{G^R_{xx}}{\omega^2} = - \frac{N}{\omega^2 - k^2 } \frac{\mathcal{B}_x }{\mathcal{A}_x} = \frac{G^R_{tt}}{k^2}  \, .
\label{eq:GreenFun}
\eeq
Evaluating this expression for the solution in Eq. \eqref{eq:ExAss}, one obtains the real part of ac resistivity of the BTZ background \cite{max-ads3}, $N \rho_{ac} = (\omega^2 - k^2)^2/(\pi T \omega^2)$. In the dc limit, the resistivity vanishes, which is expected since the background is translationally invariant. Adding the impurity (or a localized condensate) will break the translation invariance, and $\rho_{\rm dc}$ cannot vanish anymore. 

For $T < T_c$, the presence of the impurity changes the asymptotic structure of the gauge fields where the new terms do not source any new quantum field theory but are fixed by the impurity parameters. The  asymptotic expansion of $E_x$ after turning on $j_0$ is,
\bea
&& E_x \sim   \mathcal{A}'_x \ln z + \mathcal{B}'_x + E_x^{imp} \, , \nn \\
&& z E_x^{imp} = b_0 + b_1 + (b_1 + 2b_2) \ln z + b_2 (1-z) (\ln z)^2
 \, , \nn \\
&& \mathcal{B}'_x = \frac{\mu_e}{4} \left( 2 i \omega + \omega^2 \right)+ \frac{15}{8} b_2 + b_1 - b_0  - \mathcal{A}'_x \ln \Lambda \, , \nn \\
&&  \mathcal{A}'_x = \mu_e (k^2 - \omega^2) - ( b_1 +  2 b_2) \, .
\label{eq:Efield}
\eea
Here, $b_0/b_2 = 1+ (1/\kappa -1)^2 \, , \,\, b_1/b_2 = 2 ( 1/\kappa - 1) \, ,$ and  $\, b_2 = - 2 \mathcal{Q}_i \alpha^2 \, $, are the impurity parameters, and $\mu_e$ is the chemical potential corresponding to the gauge field $A_\mu$. The derivation of these coefficients is presented in Appendix \ref{sec:HydroEimp}. 

Now, we need to correctly insert the temperature scales, which can be done by understanding the marginality requirement of the Kondo deformation.
The parameter choice we made for solving $\phi(z)$ is motivated by the fact that the boundary dual operator of $\phi^2$ needs to be a marginal operator. In such a scenario, following the AdS/CFT dictionary \cite{WittMultitrace}, we can identify $\kappa$ with the double-trace coupling and $\phi$ sources a holographic superconductor on the boundary \cite{HoloSC}. In other words,  $\kappa$  plays the role of Kondo coupling in the holographic model. So far, we have been working with the dimensionless coupling $\kappa$, which we now replace with $- \kappa/ N \lambda_K$ \cite{HoloKondo}.  In the asymptotic solution of $\phi(z)$ in Eq. \eqref{eq:SolnPhi}, $\Lambda$ is an arbitrary renormalization scale which should not dictate the solution $\phi(z)$.  Hence, by demanding $\phi(z)$ remains unchanged under this rescaling $\Lambda_0 \rightarrow \Lambda$, 
the renormalized parameters are related to the bare parameters by $\kappa^{-1} + \ln \Lambda = \kappa_0^{-1} + \ln \Lambda_0$. After performing a thermal stability analysis \cite{HoloKondo} this translates into an insertion of the following temperature scales,
\beq
\frac{\kappa}{\kappa_T} = - \ln \left( \frac{T}{T_K} \right) \quad , \quad \alpha_T^2 = \frac{\alpha^2}{2 \pi T} \, ,
\label{eq:TempDep}
\eeq
where $T_K$ is defined as the temperature at which $\kappa_T$ (Kondo coupling at finite temperature) diverges. Thus, $T_K$ can be identified as the Kondo temperature. For an antiferromagnetic coupling, $\kappa < 0$, and $\kappa_T >0$. We arbitrarily fix $\kappa=-1$. All these simplifications result in [using Eqs. \eqref{eq:Efield} and \eqref{eq:TempDep} in Eq. \eqref{eq:GreenFun}]
\bea
&& - \omega \, \text{Im} \left( \frac{\mathcal{A}_x}{\mathcal{B}_x} \right) =  { ( \omega^2- k^2) - R(T) } \, , \nn \\
\text{where,}&& \quad b_1 + 2 b_2 = - 2 Q_0 \alpha_T^2 / {\kappa}_T \equiv - \mu_e R(T) \, , \nn \\
\text{and,}&& \quad R(T) = R_0  \lambda_K \ln \left( \frac{T}{T_K} \right) ,\,  R_0 = \frac{ \mu_i \expect{\mathcal{O}}^2}{N \mu_e T_c} \, .
\label{eq:R(T)}
\eea
The $\langle{\cal O}\rangle^2$ dependence of the imaginary part of the pole of the Green function is in agreement with the resonance in the spectral function found previously \cite{HoloKondoFano}.
The simplification of $R(T)$ done in the last line is performed by inserting appropriate temperature scales, as shown in Eq. \eqref{eq:TempDep}. Note, that since $\expect{\mathcal{O}} \sim O(N)$ and $\lambda_K \sim O(1/N)$, therefore $R_0 \lambda_K \sim O(N^0)$. $R(T)$ characterizes the decay width of the pole in the Green function. This decay is a characteristic of the impurity in the model, and it is this logarithm pole which causes the log-rise in the resistivity. Plugging in the above simplified expression [Eq. \eqref{eq:R(T)}] for the Green function in the Kubo formula in Eq. \eqref{eq:ResistivityForm}, we rewrite the resistivity as
\bea
\rho (\omega, k) &=&  \frac{\omega^2-k^2}{ N \omega^2} \left( \frac{\omega^2-k^2}{ \pi T}  -  R(T)  \right) 
\, , \\
\rho_{dc}(T) &=&   - \frac{ R_0}{N}  \lambda_K \, \ln \frac{T}{T_K} \, .
\eea

This is the characteristic logarithmic increase of the resistivity in the Kondo problem. The explicit $1/N$ dependence in $\rho_{dc}$ is clear from the fact that $N$ acts as a coupling constant in the action in Eq. \eqref{eq:Action}, $S \propto N$. This makes all the $n$-point functions proportional to $N$ and $\rho \propto \langle J_xJ_x \rangle^{-1} \propto 1/N$. This implies for a classical spin, $SU(N \rightarrow \infty)$, the resistiivity vanishes, correctly establishing the fact that Kondo effect is a pure quantum mechanical effect, arising in the original context from the noncommutativity of internal degrees of freedom. Also, in the absence of the screening cloud, which makes  $R_0=0$, the resistivity vanishes. If we take the limit $T \rightarrow \infty$, before setting $\omega, k \rightarrow 0$, the resistivity also vanishes. This is expected since at the UV fixed point, $\lambda_K = 0$ (hence $\kappa_T = 0$ or $T \rightarrow \infty$), the impurity and the electrons get decoupled. As a result of the negative sign in front of  $\rho_{dc}$, the log-behavior can dominate only in case of antiferromagnetic coupling ($\lambda_K > 0$). This calculation holds true for $T< T_c \sim T_K$, since in the slave fermion formulation the Kondo effect is enabled by superconductivity. For describing the physics in the temperature range $T \gg T_K$, one needs to go beyond the saddle point approximation and consider a backreacted geometry. 

It appears from our calculation that the $\ln T$ dependence of the resistivity can appear anytime there is an impurity with any internal degree of freedom (such as a spin or orbital momentum), as long as its coupling to the host matter is marginally relevant. In other words, a scalar impurity with a marginal coupling to the host fermionic matter cannot give rise to the Kondo-like effect. A trial term to see this is to consider a coupling of the form $H_K = \lambda_K |\varphi|^2 \psi^\dag \psi$, where $\varphi$ is the localized impurity field. Obviously, $\mathcal{O}^\dag = \varphi \psi^\dag$ being a fermionic operator cannot condense. This would mean that some electrons are always stuck to the impurity no matter how fast they move. Hence, this operator cannot capture a physical process. There is only one other possible scalar coupling $\mathcal{O} =  |\varphi|^2 \psi^\dag \psi $, which is also marginal. However, the scalar bulk dual of $\mathcal{O}$ cannot be charged since it is a $U(1)_\psi$ singlet.  As a result, it cannot be coupled to the probe field, giving rise to no log-correction. All of this implies that  the primary ingredient in a holographic log-rise of the resistivity is the classically marginal double-trace coupling $\mathcal{O}^\dag \mathcal{O}$, where $\mathcal{O}$ must have a global $U(1)$ symmetry in order to affect the two-point function [of the $U(1)$ current].

We briefly comment about the IR resistivity. This can be computed by constructing near-horizon solutions of the impurity fields. Considering the near-horizon solutions of the EOM in Eqs. \eqref{eq:EOMPhi} and \eqref{eq:EOMat}, owing to the boundary conditions, at the IR fixed point, one obtains $a_t(z_h)=0$, and $\phi(z_h)$ is a constant. This causes the source to the probe fields to vanish, $j_0(z_h) = 0$; hence, one cannot see any impurity effects in the resistivity. This is expected since at the IR fixed point the impurity is completely screened, restoring the translational invariance to the system. A more interesting quantity could be to compute the first leading deformation to this by flowing the theory slightly away from the IR fixed point. This can be done by constructing the leading irrelevant operator $\mathcal{O}_{\text{ir}}$, of dimension $\Delta_{\text{irr}} > 1$. In other words, one needs to solve for the leading corrections of $a_t(z)$ and $\phi(z)$ near the horizon. Hence we add a perturbation, $\int d^2x \delta(x) \lambda_{ir} \mathcal{O}_{ir}$, where the irrelevant coupling $\lambda_{ir}$ is of dimension $1-\Delta_{\rm ir}$. Using scaling arguments, it can be shown \cite{HoloKondo} that the leading correction to all the thermodynamic quantities will be power laws in $T$. Specifically, $\rho_{dc} \propto \lambda_{ir}^2 T^{2(\Delta_{ir} - 1)}$. 

\section{Conclusion}
 In a large-$N$ holographic model of the Kondo Hamiltonian, the Kondo effect can be essentially  seen as the formation of a screening cloud around the magnetic impurity, which is achieved by condensing $\mathcal{O}$ in the holographic model. This induces a charge density at the impurity site, which vanishes for $T>T_c$. The condensation is driven by a marginal coupling $\kappa$, which is dual to the Kondo coupling $\lambda_K$. Marginality of $\mathcal{O}$ causes the impurity charge density to run logarithmically in the bulk, which modifies the asymptotic solutions of the boundary probe field, giving rise to a logarithmic decay width of a pole in the Green function. This decay in the pole manifests as the log-correction to resistivity.   We emphasize here that in our holographic calculation near the strong-coupling fixed point (IR regime) the resistivity remains a power law; however, near the weak-coupling fixed point (UV limit), the resistivity scales as $\ln T$.  In a conventional perturbative approach to the Kondo model with a free electron host, one expects a logarithmic dependence of the resistivity only for $T\gg T_K$. This result cannot be anticipated to be achieved using a classical holographic setup for two reasons. First, in the $N\rightarrow \infty$ limit, one expects a sharp transition at the Kondo temperature with the condensate taking a finite value only below $T_K$. This transition is rounded with finite $1/N$ corrections \cite{AndreiLargeN} giving rise to a nonzero value of the condensate (and hence a log-rise of the resistivity) even above $T_K$. Second, the strong correlation with the host changes the applicability of the result near the Kondo regime. Indeed, for a strongly correlated electron host \cite{BickersPRB}, it was found using the Bethe ansatz \cite{weigman,andrei} that the logarithmic behavior extends to low temperatures less than or above the Kondo temperature, fully consistent with our finding. Consequently, this work certainly expands the appeal of holography as a tool for solving problems in strongly coupled electron matter.

\section*{Acknowledgments }
We thank B. Langley, and S. Balakrishnan for many helpful discussions and J. Erdmenger, A. O'Bannon, and T. Faulkner for valuable comments on the manuscript. C.S. acknowledges support from Center for Emergent Superconductivity, a DOE Energy Frontier Research Center, Grant No. DE-AC0298CH1088.  We are also thankful for the NSF Grant No. DMR-1461952 for partial funding of this project. 


\appendix
\numberwithin{equation}{section}
\section{Hydrodynamic Solution of Maxwell \ads{2}}
\label{sec:HydroE}

Instead of directly solving the EOM for $E_x$, we find it easier to solve $A_t'$ first and then obtain $E_x$ from it. The equations we use are
\begin{gather}
A_t''' + \frac{(h z)'}{h z} A_t '' + \left( \frac{h'}{z h} - \frac{1}{z^2} + \frac{p^2}{h^2} \right) A_t' = 0 
\label{eq:AtPEOM} \, , \\ 
E_x = \frac{h}{k} \left( A_t'' + \frac{1}{z} A_t' \right) \, .
\label{eq:ExEOM}
\end{gather}
We now use the hydrodynamic expansion method to solve Eq. \eqref{eq:AtPEOM}. This allows us to perturbatively understand the long-distance and low-frequency behavior of the theory at a finite temperature. First, there is a (regular) singularity at $z=1$, which can be removed by (Fuch's theorem) writing $A'_t = (1-z)^\nu F$. $\nu$ follows a quadtratic equation, and the solutions are $\nu_\pm = \pm i \omega/2$. Choosing an  infalling boundary condition, we keep $\nu_-$; hence, the differential equation obeyed by $F(z, \bs k) $ is
\begin{widetext}
\beq 
F'' + \left( \frac{1-3z^2}{z h} + \frac{i \omega (1+z)}{h} \right) F' + \frac{i \omega}{2} \left( \frac{1+2z}{z h} \right) F - \left( \frac{ \omega^2 (1+z)^2}{4 h^2}   + \frac{1+z^2}{z^2 h }\right) F - \frac{p^2}{h} F = 0 \, .
\eeq
\end{widetext}
For $\omega, k \ll 1$ we expand $F(z, \bs k) \simeq W_0(z) + \omega W_1(z) + k K_1(z) + \omega^2 W_2(z) + k^2 K_2(z) + \omega k G_1(z)$. Substituting this in the above equation and then solving the coefficient functions up to a constant factor, we get
\bea
W_0 &=& 1/z \, , \, W_1 =  \frac{ \ln(1+z) }{2z i} \quad , \quad
K_2 = - \frac{2 \ln z \ln(1-z^2) + \Li{z^2}}{4z} \, , \nn 
\\
16 z W_2 &=& 2 \Li{ \frac{1-z}{2}} + 2 \Li{\frac{1+z}{2}} + 8 \Li{z} - 8 \Li{1+z}  - 2 \ln^2(1+z) \nn \\ 
 &+& 2\ln(1-z) \left(  \ln\left(\frac{1+z}{2}\right)+ 4\ln z \right) - (8 i \pi + \ln 4) \ln(1+z) + \pi^2 + 2 \ln^2 2 \, .
\eea
All the integration constants are fixed by demanding that the higher-order coefficients $W_n$s and $K_m$s vanish at the boundary (the asymptotic form of $A'_t$ is fixed to be $1/z$), and near the horizon, they should be regular. Since the equation of $F$ is symmetric in $k \rightarrow -k$, all linear terms in $k$ such as $K_1(z)$ or $G_1(z)$ trivially become zero. Here, $\Li{z}$ is the polylog function and is defined by the power series  $\sum_{n=1}^\infty {z^n}/{n^2}$. Even though we can solve $A_t'(z)$ exactly what is more useful is its asymptotic behavior
\beq
A_t'(z, \bs k) = \frac{1}{z} + \frac{1}{2}(k^2 - \omega^2) z \ln z + \frac{1}{8} (-2 k^2 + 2i\omega + 3\omega^2) z \, .
\label{eq:AtAss}
\eeq
Combing this with Eq. \eqref{eq:ExEOM}, we obtain the solution presented in Eq. \eqref{eq:ExAss} of the main text. 
\\
\section{Hydrodynamic Solution of Impure Maxwell \ads{3}}
\label{sec:HydroEimp}

We want to solve the Maxwell equations in the presence of a source,
\beq
\frac{(j_0/z)'}{h} = A_t''' + \frac{(h z)'}{h z} A_t '' + \left( \frac{h'}{z h} - \frac{1}{z^2} + \frac{p^2}{h^2} \right) A_t' 
\, .
\label{eq:sourcedAt} 
\eeq
 In the previous section, we did so in the absence of the source; hence, we can make use of this homogeneous solution and add a particular solution to it. We begin with the following ansatz (power-series) solution:
 \begin{widetext}
\bea
A'_t(z) & = & \frac{a_0}{z} + b_0 + b_1 \ln z + b_2 (\ln z)^2 + z \left( c_0 + c_1 \ln z + c_2 (\ln z)^2 \right)  + O(z^2) \, , \\
\text{or,} \quad 
A_t(z) &=& a_1 + a_0 \ln z + z (b_0 - b_1 + 2 b_2 + (b_1 - 2 b_2) \ln z + b_2 (\ln z)^2)   \nn \\
&& + \, \frac{z^2}{4} (2c_0 - c_1 + c_2 + 2 (c_1  - c_2) \ln z + 2 c_2 (\ln z)^2) \, ; \\
\text{and} \quad k E_x(z) &=&  \frac{b_0 + b_1 + (b_1 + 2 b_2) \ln z + b_2 (\ln z)^2}{z}
 \nn \\ && 
 + \, 2 c_0 + c_1 + 2 (c_1 + c_2)   \ln z  + 2 c_2 (\ln z)^2 \, .
 \label{eq:AssEx}
\eea
\end{widetext}
The coefficients $a_i$s can be fixed from boundary conditions of $A_t$. Like before, we demand regularity at the horizon and a chemical potential for the charge density described by $A_t$ at the boundary. Hence,
\beq
a_0 = \mu_e \quad , \quad a_1 = b_1 - b_0 + 2b_2 - b_2/8  \, .
\eeq
The newly appearing coefficients in $A_t'(z)$ should be fixed by the source term. For that, we consider the asymptotic expansion of Eq. \eqref{eq:sourcedAt},
\beq
A_t''' + \frac{1}{z} A_t'' - \frac{1}{z^2} A_t' = (Q/z)' \, .
\eeq
Matching the coefficients on either side, we obtain
\begin{gather}
b_0/b_2 = 1+ (1/\kappa -1)^2 \, , \,\, b_1/b_2 = 2 (1/\kappa - 1) \,\, , \nn \\
2c_2/b_2 = - 1 \, ,  \,\, b_2 = - 2 \mathcal{Q}_i \alpha^2 \, .
\label{eq:ImpCof}
\end{gather}
Now, there remain two unfixed constants, $c_0$ and $c_1$, which can be solved from the EOM. On the boundary, one of them plays the role of the source, and the other is the corresponding vacuum expectation value. In order to determine the $c_i$ terms, we have to analytically solve the EOM in Eq. \eqref{eq:sourcedAt}. This can be solved using a trick for nonhomogeneous differential equations,\footnote{Given a generic second-order differential equation $$F''(z) + a_2(z) F'(z) + a_1(z) F(z) + a_0(z) = S(z) \, ,$$ if $u(z)$ solves the equation in the absence of the source  $S(z)$, then once can construct the full solution as $F(z) = u(z) v(z)$, where the equation for $v(z)$ is $\left(u E(z) v' \right)' = E(z) S(z)$. Here, $E(z) = u \exp\left(\int dz \, a_2(z) \right)$ appears as a factor to make the equation exact. In our case, $A_t'(z)$ is treated as $F(z)$, and for Eq. \eqref{eq:sourcedAt}, $E(z) = zhu(z)$. Alternatively, if we are looking for a series solution then it is sometimes easier just to perform a superposition of solutions with and without the source (found by matching each term in the series with the source, order by order).
} 
and we get $A_t'(z)=u(z) v(z)$ where $u(z)$ solves the homogeneous EOM in Eq. \eqref{eq:AtPEOM}, and $v(z)$ obeys 
\bea
\left( z h u^2(z)  v'(z) \right)' = z u(z) (j_0(z) /z)' \, .
\eea
The above equation being exact, one can formally integrate out $v(z)$; however, this integration may not be very tractable. Hence, instead of working with an exact expression of $u(z)$ (which is not available to us anyway), we work with its $\mathcal{O}(z)$ asymptotic solution obtained in Eq. \eqref{eq:AtAss}. Note that we have $j_0(z)$ of the order $\mathcal{O}(z)$, hence from the above equation, we can obtain a $\mathcal{O}(1)$ solution to $v(z)$, and hence a $\mathcal{O}(z)$ solution to $A_t'$. This is already sufficient since at least for our purpose we need no more than the sub-sub-leading behavior of $A_t'$. As one last thing, since at the horizon $j_0(z)$ vanishes, the leading behavior near the horizon does not change; hence, the boundary conditions remain intact. However, as we will see, for our purpose we do not need to fix these constants. Taking care of these facts and after performing an integration, we obtain
\begin{gather}
8 c_0( \bs k) = \mu_e (-2 k^2 + 2i\omega + 3\omega^2) + 4 a_1 + 2 (b_1+b_2)  \, , \nn \\
2 c_1(\bs k) = \mu_e (k^2 - \omega^2) - (b_1 + b_2) \, .
\end{gather}
This way also we obtain the same results for $b_i$s as in Eq. \eqref{eq:ImpCof}. Plugging all the coefficients in Eq. \eqref{eq:AssEx} we obtain 
\begin{gather}
k E_x  =  \frac{b_0+b_1 + (b_1 + 2b_2) \ln z + b_2 (1-z) (\ln z)^2}{z}  + \mathcal{A}'_x \ln z + \mathcal{B}'_x \, ,  \\
- \mathcal{A}'_x =  \mu_e (\omega^2 - k^2 ) +  b_1 +  2 b_2 
\, , 
\nn \\ 
\mathcal{B}'_x = a_1 + \frac{\mu_e}{4} \left( 2 i \omega + \omega^2 \right) - \mathcal{A}'_x \ln \Lambda \, .
\end{gather}

\pagebreak


\end{document}